\documentclass[]{aa}
\usepackage{natbib}
\usepackage{psfig}
\def\HH{{\rm H}_2}
\def\nH2{{\rm n}({\rm H}_2)}
\def\NH2{{\rm N}({\rm H}_2)}
\def\pccc{{\rm cm}^{-3}} \def\pcc{{\rm cm}^{-2}}

\def\Tstar#1 {\mbox{${\rm T}_{\rm #1}^*$}}
\def\Tsub#1 {\mbox{$T_{\rm #1}$}}
\def\Tk  {\Tsub K }

\def\Ta  {\Tstar A }

 \def\arcmin{\mbox{$^{\prime}$}}

\def\degr{$^{\rm o}$}
\def\p{$^+$}

\def\hcop{\mbox{{HCO\p}}}

\def\cch{\mbox{C$_2$H}}

\def\h13cop{\mbox{{H$^{13}$CO\p}}}
\def\nnhp{\mbox{N$_2$H\p}}
\def\c3h2{\mbox{C$_3$H$_2$}}
 
 \def\R0{R$_0$}

\def\ddeg{{}^\circ\kern-.1em}

\def\kms{\mbox{km\,s$^{-1}$}}

\def\E#1{\,10^{#1}}
\def\P#1,{$\nH2\Tk~=~#1\times~10^4~\pccc$~K}
\def\ec#1,#2,#3,{#1\,(#2)\E{#3}}

\def\zoph{$\zeta$ Oph}
\def\so2{SO$_2$}
\title{Comparative Chemistry of Diffuse Clouds III:  Sulfur-bearing
Molecules}

\titlerunning{Sulfur-Bearing Molecules ...}
\author{R. Lucas\inst{1}\ and H. S. Liszt\inst{2}}
\institute{Institut de Radioastronomie Millim\'etrique,
           300 Rue de la Piscine,
           F-38406 Saint Martin d'H\`eres,
           France
\and       National Radio Astronomy Observatory,
           520 Edgemont Road,
           Charlottesville, VA,
           USA 22903-2475}
\begin{document}
\date{received \today}
\offprints{H. S. Liszt}
\mail{hliszt@nrao.edu}

\abstract{
Using the Plateau de Bure Interferometer and IRAM 30m Telescope, we 
observed $\lambda$2-3mm absorption lines of CS, SO, \so2, $\HH$S 
and HCS\p\ from some of the diffuse clouds which occult our well-studied
sample of compact extragalactic mm-wave continuum sources.  Our
observations of SO, $\HH$S and HCS\p\ represent the first detections of
these species in diffuse clouds; \so2\ was not detected at all. We 
find a typical value 
N(CS) $\approx 0.5 - 2.0 \times 10^{12}~\pcc$ which is actually 
much smaller than the values derived previously, along very different 
lines of sight, from interpretation of CS J=2-1 emission lines.  CS 
forms somewhat sluggishly and is occasionally absent even in features 
with appreciable N(\hcop).  But for lines of sight where CS is found, 
N(CS)/N(\hcop) $\approx 2\pm1$ or X(CS) $\approx 4 \times10^{-9}$.  N(CS)
correlates well and varies about linearly with N(HCN) 
(N(CS)/N(HCN) = $0.7\pm0.3$) though, again, lines of sight with
appreciable N(HCN) occasionally lack CS.  CS correlates well with 
$\HH$S (N(CS)/N($\HH$S) $ = 6\pm1$) and marginally with 
SO (N(CS)/N(SO) = $1.7\pm0.8$).  For the two high column density 
features observed toward 3C111 we find N(CS)/N(HCS$^+$) = 13.3$\pm1.0$.
CS can easily be shown to form from the observed amounts of
HCS\p\ $via$ electron recombination in cool, quiescent gas but the
obvious gas-phase routes to formation of HCS\p\ fail by factors
of 25 or more.
}
\maketitle

\section {Introduction.}

This is the third in a continuing series of papers setting forth
the run of molecular abundances derived from the mm-wave absorption
spectra of diffuse clouds (A$_{\rm V} \la 1$).  For the most part the 
abundances traced in this work are unexpectedly large, and the species 
-- polyatomic -- 
are of a complexity which is unexpected for a medium threaded by the 
harsh interstellar radiation field \citep{DrdKna+89}.  For reasons 
which are not presently understood,
neither the comparatively low density of these regions nor the absence of 
shielding by dust is a substantial barrier to chemical complexity.
Many of these clouds are only barely dense enough or opaque enough to support
large abundances of molecular hydrogen and they are marked by an absence of 
appreciable mm-wave molecular emission in any species beside CO 
\citep{LisLuc00}. Even CO is occasionally absent in emission despite the 
clear presence of an absorption feature \citep{LisLuc98}.

The two earlier papers in this series concerned families of molecules
whose presence in the interstellar medium first became known when their
simplest, diatomic members (CH, CN) appeared in absorption toward 
nearby bright stars 
\footnote{
 Another possible family, starting with OH and \hcop\ and leading to
 CO, has been the subject of much of our prior work 
 \citep{LucLis96,LisLuc96,LisLuc00}.
}
In Paper I \citep{LucLis00} we considered the 
C$_{\rm n}$H$_{\rm m}$ species CH, \cch, \c3h2 {\it etc.}. We showed
that \cch\ is widespread -- detectable in every feature seen in \hcop,
with N(\cch)/N(\hcop) higher at small N(\hcop) or N($\HH$) -- and about
linearly related in column density to \c3h2.  \c3h2\ was already known to 
be ubiquitous from prior work at cm-wavelengths \citep{CoxGue+88}.

In Paper II we examined the cyanogen-bearing species CN, HCN, and HNC 
\citep{LucLis00,LisLuc00} and discovered a very tight linear proportionality 
among the column densities of all three of these species, analogous to what 
we had seen previously between for OH and \hcop.  This enabled another 
calibration of the molecular abundance scale based on the existing body of 
optical/$uv$ absorption work relating N(CN) and N($\HH$) along a few lines 
of sight.  From the fact that the CN-family abundances vary with N($\HH$) 
at nearly fixed values of the total visual extinction along those lines of 
sight, we inferred (after modelling the formation of $\HH$ in diffuse gas 
of varying number and column density) that the trace molecules increase in 
quantity and abundance even as $\HH$ itself does. 

In this work we address the sulfur-bearing molecules CS, SO, $\HH$S, HCS$^+$ 
and SO$_2$ whose presence in the interstellar medium was only found later,  
through mm-wavelength emission originating in GMC's and dark molecular clouds. 
With the possible exception of CS, the first four of these species have been 
entirely unknown in the diffuse gas until the present work, and conventional 
gas-phase ion-chemistry models predict that they should not exist in detectable 
quantities there (see Sect. 4).  
The sulfur-bearing molecules are exceptionally interesting species to detect
in the diffuse environment because models of their gas-phase chemistry in 
translucent, dark and dense clouds appear to fail in rather spectacular 
fashion;  OH and/or O$_2$ are required to be overabundant, compared to direct 
observation, by very large factors (20 - 200) in order to explain even the 
simpler species like CS and SO (see Sect. 5.1).

More generally, though it is often claimed that models of 
quiescent gas-phase ion-molecule chemistry can reproduce (or nearly so) 
the abundances of simpler group members like CH, CN, CS, or OH in diffuse
clouds  (although not CO), it usually transpires that these same models
do not come close to explaining the abundances of polyatomics like HCN, \hcop, 
HCS\p, $etc.$ from which the diatomics form {\it via} photodissociation or
dissociative recombination in diffuse gas.  Thus the ability of simple 
models to explain the various diatomics is at best somewhat accidental and 
a deeper explanation of the full panoply of observed species is needed.

The plan of the present work is  as follows.
The manner of taking and analyzing the present observations is discussed 
in Sect. 2.  The main observational results of this work are discussed in 
Sect. 3.  Sect. 4 reviews the prior history of observations of sulfur-bearing
species in the interstellar medium and Sect. 5 discusses extant models and
theory.  Sect. 6 is a brief discussion and Sect. 7 a summary.

\begin{table}
\caption[]{Background sources and profile rms}
{
\begin{tabular}{lccccc}
\hline
Source&l& b & $\sigma_{l/c}$ &$\sigma_{l/c}$&$\sigma_{l/c}$ \\
      &\degr  &  \degr  & CS & SO & $\HH$S \\
\hline
B0212+735 & 128.93  & 11.96 & 0.046& 0.021 &\\
B0355+508 & 150.38  & $-$1.60 & 0.045 &0.014 &0.110\\
B0415+379 & 161.68 & $-$8.82 & 0.029&0.028&0.055\\
B0528+134 &  191.37 &$-$11.01 & 0.032&0.018&\\
B1730$-$130&  12.03 & $+$10.81 & 0.021& 0.018 & \\
B2200+420 &  92.13 &$-$10.40 & 0.022& 0.027&0.047 \\
B2251+158&  86.11 &$-$38.18 & 0.006 & &\\
\hline
\end{tabular}}
\\
\end{table}

\section{Observations.}

Table 1 lists the background sources observed, along with their galactic 
coordinates and (for some of the transitions) the rms noise levels 
achieved in the line/continuum ratio, which is the rms error in 
optical depth in the optically-thin limit.  
$\HH$S and SO$_2$ were observed toward B0355+508, B0415+379, and
B2200+379 at the IRAM 30m telescope in 1995 January and November
using beam-switching.  CS, SO, and HCS\p\ were observed at the
Plateau de Bure interferometer in many sessions during the years
1993-1997 and the mode of observation was identical to that described 
previously by \cite{LucLis00} and in our other prior work.  The 
beam-switched observations at the 30m are straightforward because 
there is no emission of any significance in the foreground gas. All data 
were taken with a channel separation of 78.1 kHz 
and a channel width (FWHM) of 140 kHz, corresponding to velocity 
intervals which are tabulated for the various individual species.

\begin{table}
\caption[]{Species and lines observed}
{
\begin{tabular}{lcccc}
\hline
Species&line& frequency & resolution & factor$^1$ \\
      &  &  MHz  & \kms &   \\
\hline
CS & $2-1$  & 97981 & 0.428&8.06\\
SO & $3_2 - 2_1$  & 99300 & 0.423&28.6\\
SO$_2$ & $ 3_{13} -2_{02}$  & 104029 & 0.403&26.3\\
$\HH$S & $1_{10}-1_{01}$  & 168763 & 0.249&1.80\\
HCS\p & $2-1$  & 85348 & 0.492&12.9\\

\hline
\end{tabular}}
\\
$^1$ units are $10^{12}~\pcc$(\kms)$^{-1}$
\end{table}

\begin{figure*}
\psfig{figure=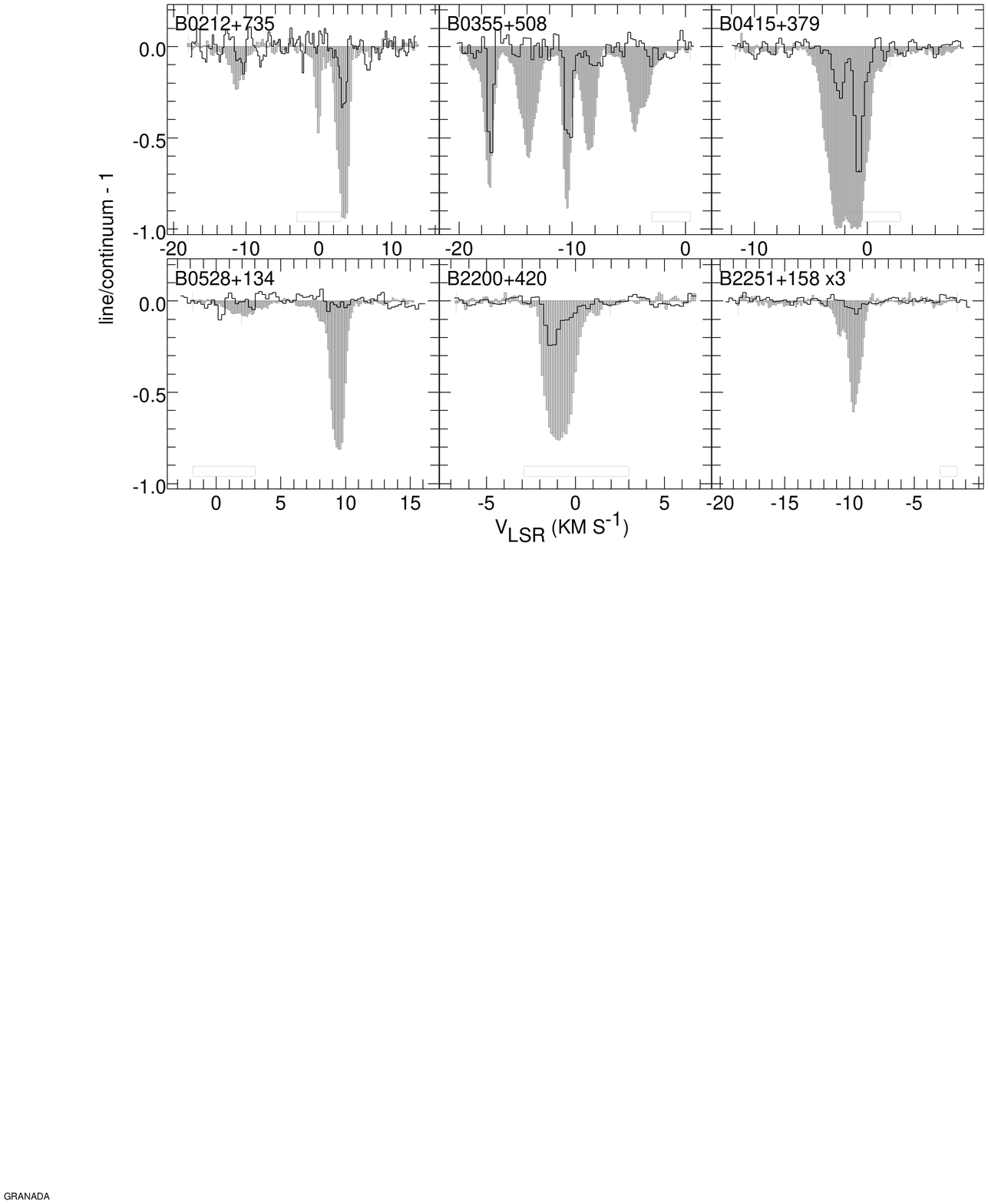,height=11.2cm}
\caption[]{CS J=2-1 (histogram) and \hcop\ J=1-0 (gray overlay) absorption 
spectra toward six sources showing all the detected CS components and the
barren spectrum toward B0528+134.}
\end{figure*}

Table 2 gives the species and transitions observed, along with the
velocity resolution (corresponding to 140 kHz) and the factor 
by which the measured integrated optical depths were scaled to 
derive the total column density.  As is usual for species having
higher dipole moments (than CO), we evaluate the rotational
partition function by assuming that the populations are in equilibrium 
with the cosmic microwave background. This follows from the diffuse nature of 
the host gas and its consequent inability to support substantial rotational
excitation in any species beside CO, even when the electron excitation
is included.  As we have noted (see, for 
example, \cite{LisLuc00}), of all the species with appreciable dipole
moments (typically 2-4 Debye), only \hcop\ has been detected in emission 
toward our sources and it is exceedingly weak even when the optical 
depth is high.  We have never succeeded in detecting CS emission
and the thermal pressures indicated by CO absorption profiles
also show that the weak-excitation limit is appropriate 
\citep{LisLuc98}.  The CS emission observed by \cite{DrdKna+89}
is consistent with this assumption, as well.

\subsection{Component fitting and profile integrals}

We did not attempt a comprehensive regimen of fitting Gaussian
components to the sulfur-molecule profiles such as we did in 
prior work, because, frankly, the data discussed here are older 
and generally of lower quality.  Instead, for all lines of sight 
except B0415+379 (where the profile integrals arise from gaussian
decomposition), we simply blocked out regions of the profiles 
according to the component structure gleaned from prior experience 
with \hcop, \cch, HCN, $etc$, and integrated over these regions.  
The basic data from which our Figures and conclusions were drawn are 
given in the Appendix.

\section{Systematics}

\subsection{N(CS) and variation across chemical families}

Figure 4 compares the variation of the CS column density (by far
the most widely-observed species) with those of representatives
of the other chemical families we have studied.  Although it is
\hcop\ whose abundance relative to $\HH$ is the most nearly
constant, and although N($\HH$) increases to the right in Fig.
5, it is not necessarily the case that variations in A$_V$ 
dominate the changes in molecular column density, as shown in
Paper II.  More likely, it is some combination of higher total
number density and extinction which drives the molecular 
fraction higher in both $\HH$ and the trace species.

Although CS is occasionally found in features of low molecular
column density, it is also occasionally absent at intermediate
values of N(\hcop), {\it etc}.  The tightest correlation, and the
most nearly linear power-law slope,  is found between N(CS) and 
N(HCN).  Again, this represents a functional dependence -- the
two species are affected in similar ways by changes in whatever
are the independent variables -- not necessarily a causal 
relationship. Typical values of N(CS) are 
N(CS) $\approx 1-2 \times 10^{12}~\pcc$, 
N(CS)/N(\hcop) $\approx 2\pm1$, averaged unweighted over all features
where both species were detected.  For those features where both CS 
and another species were detected, N(CS)/N(HCN) $= 0.7 \pm 0.3$,
N(CS)/N(\cch) $= 0.13 \pm 0.07$.

\subsection{Abundance ratios of sulfur-bearing species}

Figures 2, 3 and 5 detail the presence of SO and $\HH$S in several 
features having relatively high N(CS).  We find 
N(CS)/N($\HH$S) $ = 6\pm1$  and N(CS)/N(SO) = $1.7\pm0.8$.  Of
the seven features with N(CS) $> 10^{12}~\pcc$ in Fig. 5,
only two have N(SO)/N(CS) above unity, $\approx 1.3$.
For the two high column density features observed toward 3C111
which show HCS\p\ we find ratios N(CS)/N(HCS\p) = 
11.1$\pm2.0$ and 14.2$\pm1.3$ or a weighted mean ratio 
N(CS)/N(HCS\p)=13.3$\pm1.0$.  Abundances of the sulfur-bearing species 
all tend to increase together but only between CS and $\HH$S do the
limited data show a strong and clear correlation.

\subsection{SO$_2$}

SO$_2$ was the only molecule sought which was not detected
and its abundance is only loosely constrained by the present
work.  The upper limit on the relative abundance of SO$_2$
shown in Table 3 is typical.

\section{Prior observations and discussions of sulfur chemistry
in diffuse clouds}

\begin{figure}
\psfig{figure=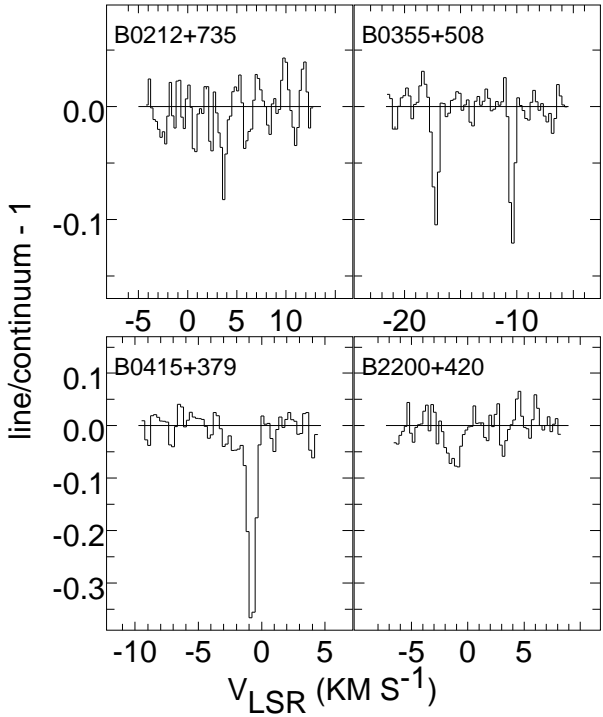,height=9.4cm}
\caption[]{Digest of detected SO absorption features.}
\end{figure}

Observationally, Snow's $Copernicus$ limits on SH toward $o$ Per,
N(SH) $< 10^{12}~\pcc$ \citep{Sno75}, and on CS toward \zoph, 
N(CS) $< 2.6 \times 10^{13}~\pcc$ \citep{Sno76},
constituted the entire body of knowledge of sulfur chemistry 
in diffuse clouds for some time.  \cite{FerRou+86} then showed that 
N(CS\p) $< 1.9\times10^{11}~\pcc$ toward \zoph, amending an earlier, 
tentative detection.  Eventually, \cite{DrdKna+89} detected very weak 
(0.01 - 0.10 K) CS J=2-1 emission toward several objects occulted by
supposedly diffuse gas (A$_V \la 1$ mag), including $o$ Per and HD210121
(T$_{\rm A}^\ast = 0.03-0.04$ K, 
N(CS) $ =  4-100 \times 10^{12}~\pcc$, 
N(CS)/N($\HH$) $ \ge 0.4 - 10 \times 10^{-8}$),
and (somewhat tentatively) \zoph\
(\Ta = 0.01 K, 
N(CS) $ =  0.7-5 \times 10^{12}~\pcc$, 
N(CS)/N($\HH$) $ = 0.15 - 1.1 \times 10^{-8}$).

The large ranges in derived CS column density in the work of
\cite{DrdKna+89} reflect a sensitivity to the assumed physical 
conditions, but the implied relative abundances are really very 
high -- in fact, quite comparable to or even larger than those 
found in dark clouds and GMC's.  
\footnote{As is evident from our work, where the typical CS column density is
N(CS) $= 1-2 \times 10^{12} ~\pcc$,  the higher CS abundances found by 
\cite{DrdKna+89} suggest that either the density was underestimated 
or that the gas sampled in emission is considerably less diffuse 
than was believed at the time.} Nonetheless, \cite{DrdKna+89} 
concluded that CS could be made in the required amount by ordinary 
gas-phase ion-molecule chemistry acting in quiescent 
diffuse/translucent gas with 
n(H) $= 200-1000~\pccc$ and $\Tk = 30-50$ K, and they 
predicted (for the line of sight toward \zoph) the diffuse-cloud 
abundances of some of the other sulfur-bearing 
species known to exist in molecular clouds, {\it i.e.} 
N(CS)/N(HCS$^+$) = 40 - 80,
N(CS)/N(SO) = 2800 - 3800, 
N(CS)/N($\HH$S) $\approx$ N(CS)/N(SO$_2$) $\approx 10^9$.
For these ratios, the observed values from the present
data (along other directions) are 13, 2, 6, and $>$ 2.

\subsection{CS emission toward \zoph}

\begin{figure}
\psfig{figure=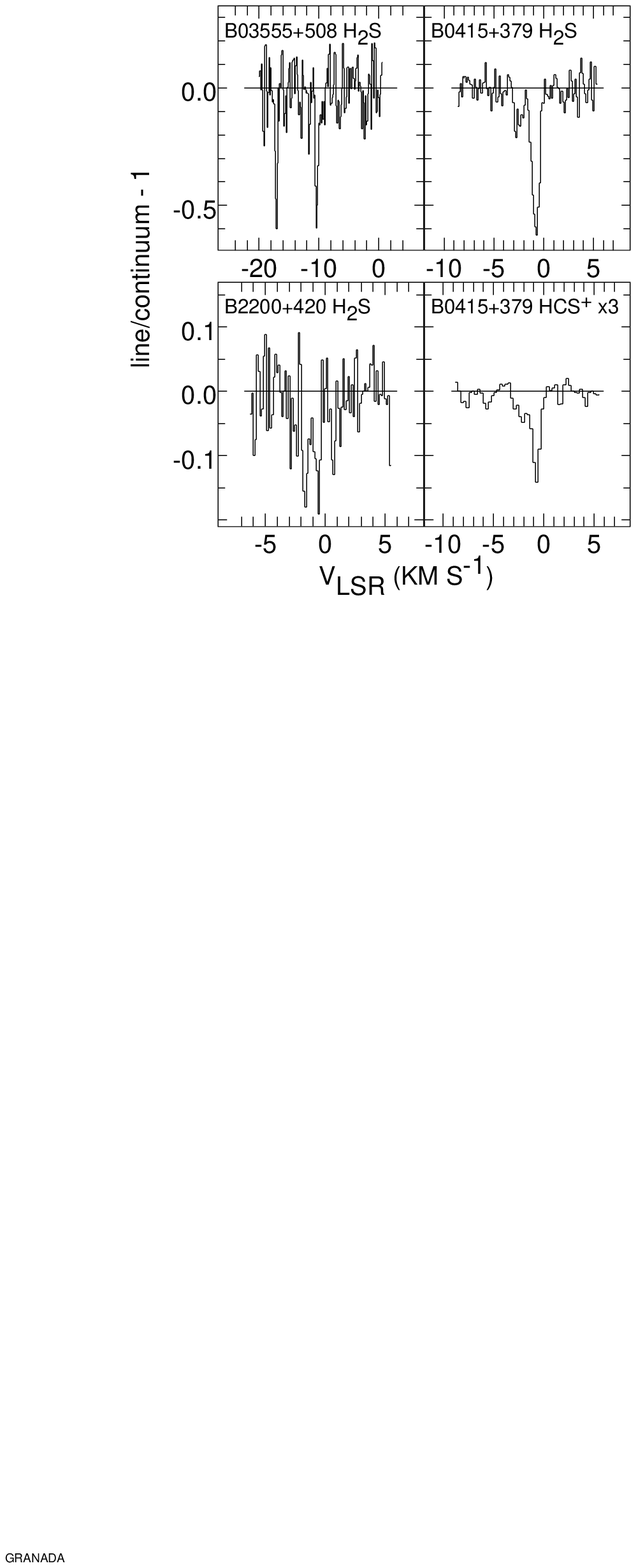,height=9.4cm}
\caption[]{Digest of detected $\HH$S and (lower right) HCS\p\ absorption 
features.  The HCS\p\ profile has been scaled upward by a factor 3.}
\end{figure}

We tried and failed (see \cite{Lis97}) to confirm the interesting 
but somewhat tentative detection of $0.013\pm0.003$ K CS J=2-1 emission 
at the somewhat unusual velocity of -1.29 \kms\ toward \zoph\ by 
\cite{DrdKna+89}.  Our spectra of 
\hcop\ and CS at two positions are shown in Fig. 6.  At the 
\hcop\ emission peak 30\arcmin\ South of the star, CS emission
is at least an order of magnitude weaker than that of \hcop.
If this proportionality were preserved, CS emission toward the star
would be six times weaker than claimed by \cite{DrdKna+89}.
Toward the star, the $1\sigma$ rms fluctuation in integrated CS 
line intensity in our data is 0.005 K \kms, but there is no signal.
So the $4\sigma$ integrated intensity detection quoted by 
\cite{DrdKna+89}, I$_{\rm CS} = 0.020 \pm0.005$ K \kms,
is a $4\sigma$ upper limit in our data.

\section{Models for sulfur chemistry} 

As is usual for the results presented in this series of papers, the 
relative abundances are similar to those believed to obtain in dark clouds 
(Table 3).   The theoretical sulfur chemistry of dark and dense clouds 
is generally (but not obviously correctly) considered to be on a firmer 
footing than that for diffuse gas and is discussed first in this Section.

\subsection{Dark gas}

Sulfur chemistry is conditioned on the inability of all species 
{SH$_{n}$}\p\ (n = 0,1...) to protonate in reaction with $\HH$ at
low temperatures (i.e. S\p\ + $\HH +9860 K \rightarrow$ SH\p\ + H): 
these reactions are all much more endothermic even than, say, 
C\p\ + $\HH + 4640 K \rightarrow$ CH\p\ + H, and CH\p\ + $\HH$ is 
exothermic. Because of this, quiescent dark and dense cloud 
gas-phase ion-molecule chemistry is usually understood as making 
SO $via$ the reaction of S with OH and/or O$_2$,  S+OX $\rightarrow$ 
SO + X, while CS forms partly from C + SO $\rightarrow$ CS + O 
(an important sink for SO) and partly from another chain initiated 
by S\p, such as 
S\p + CH $\rightarrow$ CS\p + H \citep{OppDal74,NilHja+00}. CS, 
in its turn, is destroyed in the reaction CS + O $\rightarrow$ CO + S.
 
The importance of reactions with free atomic oxygen and carbon
renders the CS/SO ratio in dark gas highly dependent on the 
elemental abundance ratio of O and C, and also to the extent to which 
these are converted to CO.  A  higher O/C ratio leaves more free 
gas-phase atomic oxygen (after CO is formed) resulting in higher 
abundances of OH and O$_2$, more rapid SO formation and CS destruction
and a larger SO/CS ratio (see Fig. 3c of  \cite{GerFal+97} which
also includes the data of \cite{Tur95,Tur96}).

\begin{figure*}
\psfig{figure=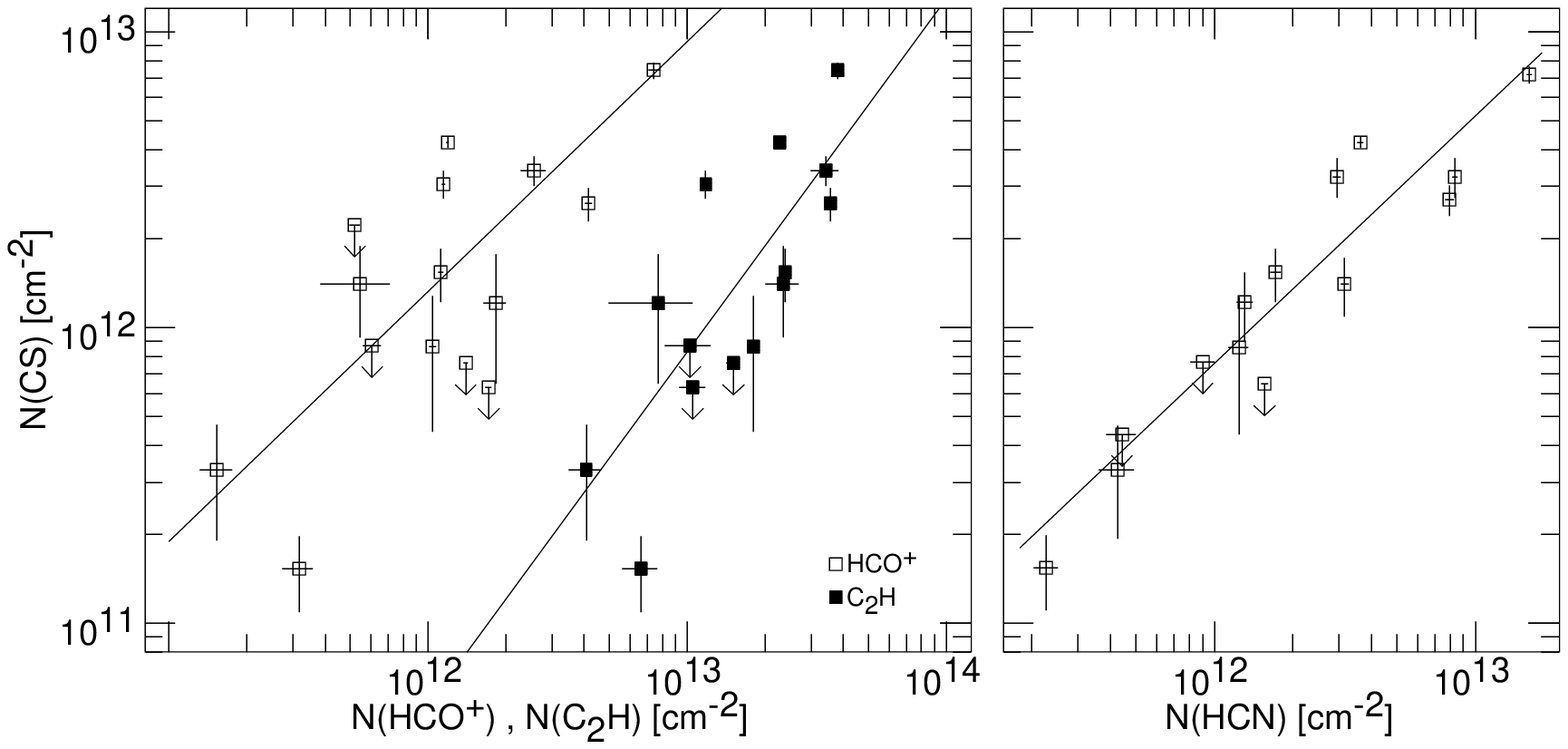,height=8.3cm}
\caption[]{Variation of N(CS) with N(\hcop), N(\cch), and N(HCN).
The power-law slopes for lines of sight with detections, 
is $0.84\pm0.20$ for CS and \hcop; 1.2$\pm0.3$ for CS and
\cch\ and 0.84$\pm$0.09 for CS and HCN.  Also for lines
of sight with detections, we have $<$CS/\hcop$>$ = 1.57$\pm0.97$, 
$<$CS/\cch$>$=0.11$\pm$0.07 and $<$CS/HCN$>$=0.72$\pm$0.28.}
\end{figure*}

SO also combines with O and OH to form SO$_2$ so the O/C ratio affects
the balance between SO and SO$_2$.  In time-dependent models it is
seen that SO$_2$ is one of the most extreme of the ``late-type'' species,
which appear in substantial quantities late in the chemical evolution
of an initially diffuse gas.  In this regard it is similar to NH$_3$
and to \nnhp \citep{WatCha85}.  In general, the amounts of OH and/or 
O$_2$ which are needed to reproduce the observed quantities of 
sulfur-bearing species in dark and dense clouds are far higher than 
those which are actually observed.  When gas-grain coupling is considered, 
high OH abundances lead to a late-time ``accretion catastrophe'' 
\citep{ChaRod+01} whereby sulfur is lost to the gas-phase entirely.

\cite{NilHja+00} have modelled the gas-phase ion-molecule chemistry in dense 
clouds and find that the very appreciable variations in mean SO/CS ratio 
between sources can be explained by factor two variations in the O/C ratio 
in the gas; spatial variations in the SO/CS ratio across individual sources 
are then to be understood in terms of density structure and other local 
behaviour.  They noted that their models entail overproduction of O$_2$ 
by very large factors as compared to existing observations, and
suggested that future searches for the elusive O$_2$ molecule might best 
be focused on lines of sight where the problems of sulfur chemistry were 
most severe in this regard.  Earlier, \cite{Tur95,Tur96} encountered similar
difficulties in attempting to account for the abundances of sulfur-bearing 
species in translucent clouds, where the required OH fractional abundance 
was found to be $2\times10^{-5}$, again much higher than the values
typically seen at  A$_V < 7$ mag, which are (with little variation)
about $10^{-7}$ \citep{Cru79,TurHei74}.

\begin{table}
\caption[]{Abundances relative to HCO\p}
{
\begin{tabular}{lccc}
\hline
Species&This work& TMC-1&L134N \\
\hline
\hcop&1.0&1.0&1.0 \\
CS     &1.6&1.3 & 0.13     \\
SO     &0.9&0.63 & 2.5    \\
SO$_2$ &$<$0.5&$<0.13$&0.5  \\
$\HH$S &0.27&0.09&0.10  \\
HCS\p  &0.13&0.08 &0.008 \\
HCN    &2.3 &2.5 &0.5 \\
\cch&13&6-12&$<6$  \\
\hline
\end{tabular}}
\\
``This work'' represents this work and data from Papers 
1 and 2 for \cch\ and HCN.  Results for TMC-1 are taken from 
\cite{OhiIrv+92} and (for $\HH$S) \cite{MinIrv+89}; L134N data 
are from \cite{OhiIrv+92}.  The abundance of \hcop\ relative to 
$\HH$ is given by \cite{OhiIrv+92} as $0.8 \times 10^{-8}$ for 
either dark cloud while a somewhat lower value $0.2 \times 10^{-8}$ 
is indicated for diffuse gas \citep{LisLuc00}.
\end{table}

\subsection{Diffuse gas}

Early-on, \cite{DulMil+80} noted that SO and SH (but not $\HH$S) might be 
formed in diffuse clouds if the radiative association of S\p\ and $\HH$ 
occurred sufficiently rapidly.  They hypothesized that the presence of 
sulfur-bearing compounds in diffuse gas, if demonstrated, might then 
point either to the efficacy of this reaction or to the importance of 
grain chemistry and gas-grain coupling.
Slightly later, but at a time when sulfur-bearing species had still 
not yet been observed in diffuse clouds, \cite{PinRou+86} showed that 
high abundances of CS, N(CS) $\ge 10^{12}~\pcc$, or
 N(CS)/N(HCS$^+$) and N(CS)/N($\HH$S) ratios of order unity, 
could be produced in the same strong (10 \kms) MHD shocks which 
seemed required to reproduce N(CH\p).  
They suggested that SH\p\ might be observable. Similar effects
enhancing the abundance of sulfur-bearing molecules in shocked
dense gas were shown by \cite{Mit84}. 

\cite{DrdKna+89} criticized the prior conclusions of both \cite{DulMil+80}
and \cite{PinRou+86}, arguing that the abundances predicted by the
shock and grain models would not have been any higher than those of
steady-state gas-phase chemistry, had the interstellar photoionization 
rates had not been underestimated.  They modelled quiescent gas-phase 
sulfur chemistry, as cited here in Sect. 4, and discussed the formation 
of CS (whose emission they observed) $via$ the interactions of S\p\ with 
CH and C$_2$ (whose abundances are well-determined) to form CS\p , 
HCS\p\ and CS (via the recombination of HCS\p).  They concluded
that the formation of CS from HCS\p\ could be modelled if 
some favorable adjustments to the CS\p\ formation rates and
CS photodestruction rates were made.

It is a fairly straightforward excercise to show that the relative 
abundance of HCS\p\ seen in this work (Table 3 with 
X(\hcop) $\equiv$ N(\hcop)/N($\HH$) $= 2\times10^{-9}$) easily 
suffices to reproduce the observed CS column densities without any of 
the favorable adjustments to rate constants and photo-dissociation 
rates discussed by \cite{DrdKna+89}.  That is, for a diffuse gas having 
an electron fraction X(e) $ = 2\times10^{-4}$ and using the standard CS 
photodestruction ($10^{-9}$ s$^{-1}$) and HCS\p\ recombination rates 
found in the UMIST reaction database, we find 
X(CS) $=  1.4\times10^{-11}$ n($\HH$)$(30/\Tk)^{0.75}$ in free space.

\begin{figure}
\psfig{figure=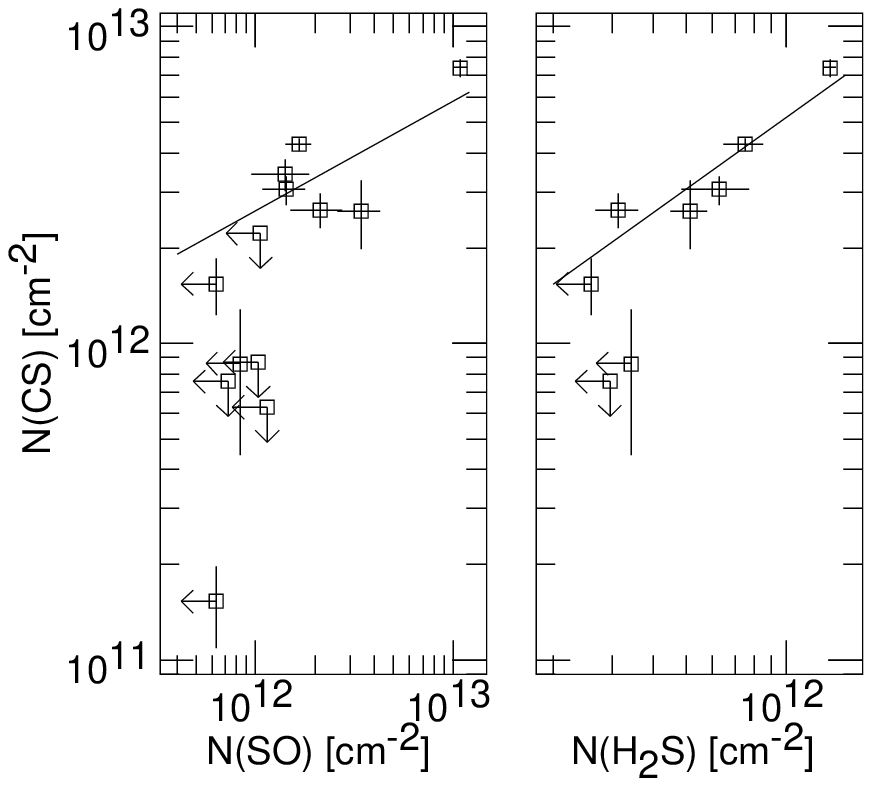,height=6.7cm}
\caption[]{Variation of N(CS) with N(SO), N($\HH$S).}
\end{figure}

However, the converse is not true; the observed diffuse gas abundances
of the species supposedly contributing to the formation of HCS\p\ 
(X(S\p) $ = 3\times 10^{-5}$; X(CH) $= 4\times 10^{-8}$; X(C$_2$) 
$= 2.4\times10^{-8}$) fall a factor of 10-20 short using quiscent gas-phase 
chemical rates (the situation is far less favorable for the other species,
as mentioned in Sect. 4).  This 
situation is essentially identical to that for CO, where the observed amounts 
of \hcop\ can easily recombine under typical quiescent conditions to 
reproduce N(CO), but the most likely progenitors of \hcop\ (OH and C\p),
both having measured abundances in at least some cases, fall similarly
short of explaining the required column density N(\hcop) 
\citep{LisLuc00}.  In principle this line of argument could be extended 
to the CN-HCN-HNC family but the permanent dipole moment of HCNH\p\ is 
small and the experiment is somewhat challenging for current instruments.

\section{Discussion}

Table 3 presents a comparison of relative abundances between the diffuse 
gas observed here and representative results for the well-studied dark 
clouds TMC-1 and L134N.  It should be noted that there are large variations 
in the CS/SO ratio between and within individual dense sources observed 
in emission (including these two dark clouds) and that TMC-1 and L134N,
whose CS/SO ratios differ by about a factor 40, represent the extremes of 
the observed range \citep{NilHja+00,GerFal+97}.  The abundances of CS and 
\hcop\ relative to $\HH$ in diffuse clouds are about 25\%-50\% that 
inferred for TMC-1.

The abundance ratios relative to \hcop\ are essentially identical for 
our sample of diffuse clouds and the representative conditions quoted
for TMC-1, with the possible exception of an overabundance of $\HH$S
in our objects.
For the dense clouds and GMC studied by \cite{NilHja+00}, the typical SO/CS 
ratio is 0.2 (i.e. it is TMC-1-like) with only one-fifth at or above unity;
this is just what is seen in diffuse clouds as well.  But
for translucent and dark clouds, ratios SO/CS $<$ 1 are actually the 
exception as detailed in Fig. 3c of \cite{GerFal+97} which includes the 
data collected by \cite{Tur95,Tur96}.

\begin{table}
\caption[]{Elemental and Molecular abundance ratios}
{
\begin{tabular}{lcccc}
\hline
Ratio &Solar& \zoph & Species & Diffuse  \\
\hline
C/O    &  0.50   &  0.50   & CS/SO    & $1.7\pm0.8$  \\
       &         &        & CH/OH    & 0.48,0.52,0.24  \\ 
C/N    &  3.80   &  1.74  & CH/NH    & 22,28  \\
       &         &        & C$_2$/CN &  9  \\
       &         &        & \cch/HCN & 5   \\
S/O    &  0.025  &  0.10   & HCS\p/\hcop\ & 0.055,0.070  \\    
S/N    &  0.20   &  0.36  & CS/CN &$0.1\pm0.05$   \\
\hline
\end{tabular}}
\\
\zoph\ elemental abundances from Savage and Sembach (1996) \\
OH and CH abundances from \cite{BlaVan86} \\
NH from \cite{CraWil97} and \cite{MeyRot91} \\
C$_2$/CN from \cite{FedStr+94}, see Paper II, Fig. 3 \\
\end{table}

It is now possible to make comparisons of the abundances of several
species which differ in structure by the replacement of one atom
with that of a different element, as for example HCS\p\ and HCO\p, 
or CS and SO, {\it etc.}.  Table 4 points out that these abundance 
ratios are not very different from that of the elements which are 
exchanged, the outstanding exception being the CH/NH ratio which is 
substantially greater than either C$_2$/CN  or \cch/HCN.

\section{Summary}

A full complement of sulfur-bearing molecules may be found in diffuse or
low-extinction gas using mm-wave absorption techniques, including CS, SO, 
$\HH$S and  HCS\p.   Only the first of these might have been seen prior to 
our work; we actually find much lower CS abundances than were derived earlier 
from observations of weak J=2-1 emission in objects purported to have A$_V$ 
of order unity. 
The observed high abundances of SO, $\HH$S and HCS\p\ cannot be reproduced 
by any gas-phase ion-molecule chemistry in quiescent diffuse gas, or perhaps 
even in much denser dark and translucent gas as well:  there is a general 
consensus that the relative abundances of OH and/or O$_2$ required to
explain the sulfur chemistry of dark gas are grossly out of line with 
what is actually observed.  

The  column density of HCS\p\ detected along one line of sight is 
entirely sufficient to explain the observed CS column density $via$ 
ordinary dissociative recombination at the standard (though very
uncertain) free-space CS photodissociation rate. This is analogous to 
the situation for \hcop\ and CO, which requires a somewhat more elaborate 
calculation of the photodissociation rate but for which there is much
more data.  In both cases, the likely precusors of the ion have relatively 
well-determined abundances and the inferred formation rate is 
1-2 orders of magnitude too small under quiescent conditions.

The abundances of the sulfur-bearing molecules seem consistent with the 
predictions of MHD shock models but the kinematics of shock models are 
never evident in our line profiles for any species.  In general, it
is clear that some mechanism is putting energy into either the gas or
the reactions; macroscopic shock models do this explicitly but not acceptably.

The next paper in this series will discuss old optical and new microwave
measurements of the CH radical.

\begin{acknowledgements}

\begin{figure}
\psfig{figure=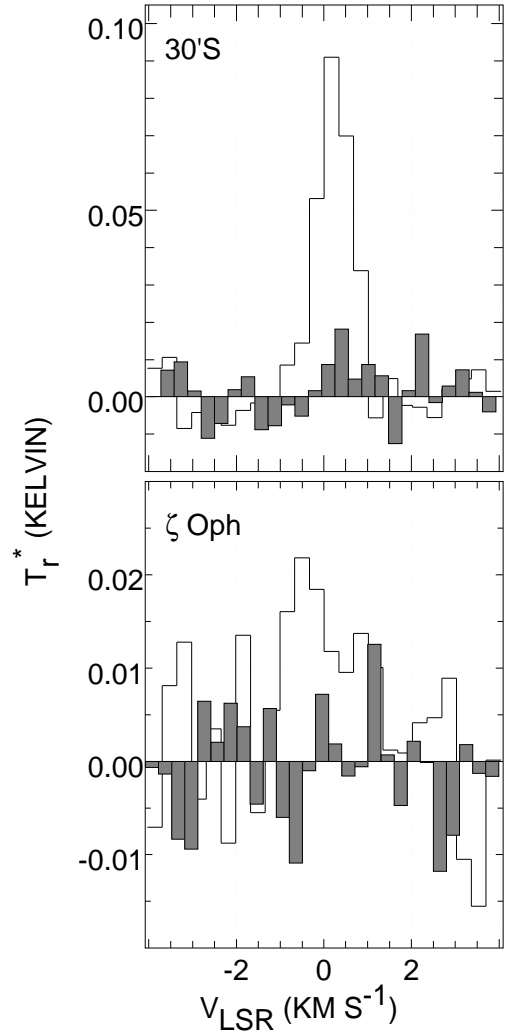,height=13.7cm}
\caption[]{CS J=2-1 (shaded) and \hcop\ (1-0) emission profiles 
toward and 30' South of \zoph.}
\end{figure}

The National Radio Astronomy Observatory is operated by AUI, Inc. under a
cooperative agreement with the US National Science Foundation.  IRAM is
operated by CNRS (France), the MPG (Germany) and the IGN (Spain). We owe the
staff at IRAM (Grenoble) and the Plateau de Bure our thanks for their 
assistance in taking the data.   We thank the referee, C. M. Walmsley,
for helpful comments. 

\end{acknowledgements}

\bibliographystyle{apj}
\bibliography{mnemonic,absorption}

\begin{appendix}
\section{Line profile integrals}
\begin{table*}
\caption[]{Line profile (optical depth) integrals}
{
\begin{tabular}{lccccc}
\hline

Source    & V    &\hcop   & \cch   &       HCN   & CS  \\
          & \kms & \kms   & \kms   & \kms        & \kms \\
\hline
B0212& -8      &0.594(0.046)& 0.378(0.077)&  0.234(0.031)&$<$ 0.11$^1$ \\ 
         &  0     &0.509(0.026)& 0.177(0.056)&              &$<$ 0.28 \\
         &  3     &2.500(0.281)& 1.259(0.150)&  4.406(0.107)&0.42(0.05)\\
B0355& -17      &1.120(0.041)& 0.434(0.022)& 1.555(0.057) &0.38(0.04) \\
         & -14      &1.367(0.028)& 0.554(0.031)& 0.477(0.051) & $<$0.09  \\
         & -10     &1.161(0.016)& 0.837(0..026)& 1.915(0.049)&0.53(0.02) \\
         & -8      &1.089(0.025)& 0.880(0.032)& 0.907(0.045) &0.19(0.04) \\
         & -3      &1.016(0.030)& 0.662(0.035)& 0.655(0.056) &0.11(0.05) \\
B0415& -2      &4.060(0.14)& 1.307(0.026)& 4.16(0.174)  &0.33(0.04) \\
         & -1      &7.27(0.37)& 1.402(0.02)& 8.532(0.212)   &0.92(0.06) \\
B0528&  10   &1.674(0.046)& 0.387(0.046)& 0.823(0.027)& $<$0.08 \\
B1730& all      &0.150(0.022)& 0.150(0.022)& 0.225(0.035)&0.04(0.02)\\
B2200&  -2     &0.533(0.16)& 0.864(0.13)& 1.67(0.063)   &0.18(0.06) \\
     &  -1     &1.78(0.18)& 0.285(0.10)& 0.690(0.050)   &0.15(0.07) \\
B2251& all      &0.31(0.042)& 0.244(0.037)& 0.120(0.013) &0.02(0.006) \\

\hline
\end{tabular}}
\\
$^1$ upper limits are 2$\sigma$ \\
\end{table*}

\begin{table*}
\caption[]{Line profile (optical depth) integrals}
{
\begin{tabular}{lccccc}
\hline

Source    &V     & SO   & SO$_2$ &  H$_2$S & HCS$^+$ \\
          & \kms & \kms & \kms   & \kms    & \kms \\
\hline
B0212    & -8     &$<$ 0.036  &          &            & \\ 
         &  0     & $<$0.037   &          &            &  \\
         &  3     & 0.049(0.016)&          &            &  \\
B0355& -17  & 0.050(0.013)&$<$0.066  &0.349(0.081)&  \\
     & -14  & $<$0.025    &$<$0.066  &$<$0.164    &  \\
     & -10  &0.058(0.009) &$<$0.045  &0.417(0.057)&  \\
     & -8   &0.022(0.039) &$<$0.058  &$<$0.144    &  \\
     & -3   &0.0292(0.052)&$<$0.074  &$<$0.190    &  \\
B0415& -2   &0.074(0.022) &$<$0.054  &0.174(0.026)&0.0185(0.0024)\\
     & -1   &0.381(0.030) &$<$0.062  &0.750(0.035)&0.0405(0.0025)\\
B0528&  10  &$<$0.040     &          &            &  \\
B2200& -2     &         &$<$ 0.024 &       &  \\
     &  -1    &         &$<$ 0.025 &        &  \\
     &  all & 0.12$\pm$0.03 &$<$ 0.035 &0.286(0.036)       &  \\
B2251& all  &$<$ 0.022    &          &            &  \\

\hline
\end{tabular}}
\\
\end{table*}

\end{appendix}

\end{document}